\documentclass[prb,aps,twocolumn,superscriptaddress]{revtex4}
\usepackage{graphicx}
\usepackage{amssymb}
\usepackage{bm}

\begin{document}

\author{V. M. Axt}
 \affiliation{Institut f{\"u}r Festk{\"o}rpertheorie,
  Westf{\"a}lische Wilhelms-Universit{\"a}t, 48149 M{\"u}nster, Germany}
\author{P. Machnikowski}
 \email{Pawel.Machnikowski@pwr.wroc.pl}
 \affiliation{Institut f{\"u}r Festk{\"o}rpertheorie,
 Westf{\"a}lische Wilhelms-Universit{\"a}t, 48149 M{\"u}nster, Germany}
 \affiliation{Institute of Physics, Wroc{\l}aw University of
 Technology, 50-370 Wroc{\l}aw, Poland}
\author{T. Kuhn}
 \affiliation{Institut f{\"u}r Festk{\"o}rpertheorie,
  Westf{\"a}lische Wilhelms-Universit{\"a}t, 48149 M{\"u}nster, Germany}

\title{Reducing decoherence of the confined exciton state in a quantum dot
by pulse-sequence control}

\begin{abstract}
We study the phonon-induced dephasing of the exciton state in a
quantum dot excited by a sequence of ultra-short pulses. We show
that the multiple-pulse control leads to a considerable
improvement of the coherence of the optically excited state. For a
fixed control time window, the optimized pulsed control often
leads to a higher degree of coherence than the control by
a smooth single Gaussian pulse.
The reduction of dephasing is considerable already for 2-3 pulses.
\end{abstract}


\maketitle

\section{Introduction}

Modern semiconductor technology offers a range of ways to produce
artificial systems where a small number of charges, confined in
all three spatial dimensions, form atomic-like systems. These
quantum dots \cite{woggon97,jacak98a} (QDs) allow one to optically
control the quantum states in a way typical for atomic systems
\cite{bayer00,zrenner02,bonadeo98,stievater01,borri02a,kamada01,htoon02,li03},
but they are still semiconductor charge devices, opening the
possibility of integration into future nanoelectronic and quantum
electronic devices \cite{bimberg99,jacak04a}.

The recent progress in ultrafast spectroscopy of semiconductor
systems \cite{axt04} made it possible to control and probe the
quantum states of carriers confined in a QD on femtosecond time
scales. However, any interaction of the confined carriers with the
external driving fields not only leads to the desired quantum
transitions but also can induce some unwanted ones. If both the
desired and unwanted transitions have a discrete nature (e.g., the
exciton vs. biexciton transition in a QD) the latter may be
suppressed by a suitable choice of control pulses
\cite{chen01,piermarocchi02}. However, since the QDs are embedded
in the macroscopic crystal, an important role is played here by
phonon-assisted transitions which inherit the continuous nature
from the phonon states. The presence of a continuum in the system
state obtained after optical excitation leads to irreversibility
of the subsequent kinetics and to dephasing of the quantum
states\cite{besombes01,krummheuer02,vagov02a,vagov03,jacak03b}
even in the absence of real transitions (energy relaxation). In
the latter case these processes are known as pure dephasing.

Effects of pure dephasing  may be observed in measurements of the
optical polarization \cite{krummheuer02,vagov02a,vagov03} or
exciton occupation in a QD
\cite{vagov02a,forstner03,machnikowski04b}. The qualitative
agreement between theoretical predictions based on the
phonon-induced pure dephasing model \cite{vagov03} and
four-wave-mixing data \cite{borri01} confirms that these
phonon-related processes dominate the dephasing in the
sub-picosecond resonant driving regime. Recently, it has been
shown that the initial decoherence in QDs with strong electronic
confinement as described by the pure dephasing model is also in
quantitative agreement with corresponding measurements
\cite{vagov04}. It turns out that even at low temperatures ($\sim$
5 K) typically about 20 \% of the coherence is lost already in the
initial phase of the dynamics within a few picoseconds. Obviously,
this rapid decoherence is not negligible and can potentially
become an obstacle for many proposed device applications
\cite{biolatti00,derinaldis02}.

The question arises whether it is possible to reduce the
phonon-assisted component of the final state by some pulse shaping
techniques, e.g., by using sequences of pulses, in analogy to the
discrete transition case \cite{chen01,piermarocchi02}. It is clear
that the continuum nature of these unwanted transitions makes this
task much more demanding. One obvious way of excluding the
phonon-assisted transitions is to make the control pulse
spectrally selective by extending the pulse durations. This is,
however, disadvantageous when other decoherence processes are
present that decrease the coherence properties of the final state
if the control operation takes too long and lead to a trade-off
situation \cite{alicki04}. For practical purposes it is therefore
essential to minimize the phonon-related pure dephasing with the
constraint that the time window for the control action is limited.

In this paper we propose a solution to this optimization problem
in terms of a control sequence composed of a series of ultrashort
(but finite) pulses. The advantage of this approach is that it is
simple both in theoretical formulation and experimental
realization, as it does not require the technical ability to
obtain arbitrary pulse shapes. Moreover, sequences of temporally
broadened pulses have simple spectral features, even in the
nonlinear regime, which lead to a transparent interpretation of
the resulting dephasing effect.

On the other hand, in spite of the apparently limited class of pulse
shapes considered, the proposed approach turns out to be unexpectedly
efficient. Already for a few pulses the degree of decoherence falls
down considerably, while a further increase of the pulse number leads
only to negligible improvement. We shall show also that achieving by a
single Gaussian pulse the same dephasing as that resulting from a
sequence of a few pulses often requires much longer control
times. Thus, the quality of coherent optical control of the excitonic
system can indeed be increased by simple means, using series of
phase-locked laser pulses, without the need to generate pulses of
arbitrary shape.

The paper is organized as follows. In the next section we present
the model of the carrier-phonon system. In Sec.~\ref{sec:delta} we
discuss the limiting case of infinitely short pulses which can be
treated analytically. In order to demonstrate the reduction of
decoherence by multiple-pulse control we compare the coherence
losses occurring after a single pulse excitation with a given
pulse area with those occurring after a pulse sequence with the
same total pulse area. Next, in Sec.~\ref{sec:formal}, we present
a perturbative method for determining the coherence loss for
arbitrarily shaped pulses and apply it to the case of
multiple-pulse sequences. The loss of coherence after a series of
driving pulses with fixed finite pulse durations and optimized
intensities is discussed in Sec.~\ref{sec:sequence}. Physically,
the duration of the pulses is limited to the lower end by the
requirement to spectrally avoid higher energetic excitations which
would be a source of further undesired dynamics and which are not
accounted for in our model. In Sec.~\ref{sec:extend} we extend our
material model and subsequently include LO phonons and higher
excited  electronic states. Explicitly including these excitations
in our optimization procedure allows us to use shorter pulses as
building blocks of our pulse sequences. The undesired dynamics
resulting from excitations of these higher energetic states is now
avoided by optimizing the pulse sequence. Finally
Sec.~\ref{sec:conc} concludes the paper.

\section{The model}
\label{sec:model}

The environment-induced process of loosing the phase relations
between different components of quantum superpositions may be
observed experimentally in a number of ways. One of its
manifestations is the decay of the optical polarization excited by
a certain laser pulse or pulse sequence resonantly coupled to the
ground-state excitonic transition in a quantum dot. The impact of
this dephasing process was studied in the case of linear
\cite{krummheuer02,besombes01} and nonlinear
\cite{vagov02a,vagov03} optical experiments.

Here, we consider the usual model of carrier-phonon interaction
involving the two states of the carrier subsystem: 
the empty dot, $|0\rangle$, and the one
exciton state, $|1\rangle$, the latter being coupled to phonon modes.
The wave functions will be modeled by anisotropic Gaussians. We
assume a simple product (Hartree) form of the exciton wave function
with the hole component shrunk in-plane due to Coulomb
interactions (this is consistent with previous results that were obtained
by full numerical diagonalization methods \cite{jacak03b}).

Thus, in the rotating wave approximation, the Hamiltonian reads
(in the rotating frame)
\begin{equation}\label{ham}
    H  =
    \frac{1}{2}f(t)(|0\rangle\!\langle 1|+|1\rangle\!\langle 0|)
    +|1\rangle\!\langle 1|
        \sum_{\bm{k}}F_{\bm{k}}(b_{\bm{k}}+b_{-\bm{k}}^{\dag}),
\end{equation}
where $f(t)$ is the pulse envelope (which will be assumed real),
$b_{\bm{k}}$ are the phonon annihilation operators for the mode
$\bm{k}$ (including implicitly also the branch index) and
$F_{\bm{k}}$ are phonon coupling constants for the ground exciton
state. In the present paper the deformation potential coupling to
the longitudinal acoustical (LA) phonons and the Fr{\"o}hlich
coupling to the longitudinal optical (LO) phonons will be
considered, with the corresponding coupling constants given by
\cite{mahan00}
\begin{equation}\label{F-LA}
    F_{\bm{k}}=\sqrt{\frac{\hbar k}{2\rho v c_{\mathrm{l}}}}
    \left[ \sigma_{\mathrm{e}}{\cal F}_{\mathrm{e}}(\bm{k})
        - \sigma_{\mathrm{h}}{\cal F}_{\mathrm{h}}(\bm{k})\right]\;\;\;
        (\mathrm{LA}),
\end{equation}
and
\begin{equation}\label{F-LO}
    F_{\bm{k}}=-\frac{e}{k}\sqrt{\frac{\hbar\Omega}%
        {2v\epsilon_{0}\tilde{\epsilon}}}
        \left[ {\cal F}_{\mathrm{e}}(\bm{k})
        - {\cal F}_{\mathrm{h}}(\bm{k})\right]\;\;\;
        (\mathrm{LO}),
\end{equation}
where $\rho$ is the crystal density, $c_{\mathrm{l}}$ is the
longitudinal sound speed, $v$ is the normalization volume of the phonon modes, 
$e$ is the elementary charge, $\epsilon_{0}$ is the vacuum dielectric
constant,
$\tilde{\epsilon}=(1/\epsilon_{\infty}-1/\epsilon_{\mathrm{s}})^{-1}$
is the effective dielectric constant, where
$\epsilon_{\mathrm{s}}$ and $\epsilon_{\infty}$ are the static and
high-frequency dielectric constants, respectively, $\Omega$ is
the LO phonon frequency and 
\begin{displaymath}
{\cal F}_{\mathrm{e,h}}(\bm{k})=\int d^{3}r\Psi_{\mathrm{e,h}}^{*}(\bm{r})
e^{i\bm{k}\cdot\bm{r}}\Psi_{\mathrm{e,h}}(\bm{r}),
\end{displaymath}
where $\Psi_{\mathrm{e,h}}(\bm{r})$ are the electron and hole
wave functions. Since the electron and hole are assumed
to overlap, the piezoelectric coupling will be
neglected\cite{krummheuer02}. The parameters used in the
computations are shown in Tab.~\ref{tab:param}.

\begin{table}
\begin{tabular}{lll}
\hline
Effective dielectric constant & $\tilde{\epsilon}$ & 62.6 \\
Longitudinal sound speed & $c_{\mathrm{l}}$ & 5600 m/s \\
LO phonon frequency & $\Omega$ & 54 ps$^{-1}$ \\
Deformation potential  & & \\
\hspace{0.7em} electrons & $\sigma_{\mathrm{e}}$ & $-8.0$ eV \\
\hspace{0.7em} holes & $\sigma_{\mathrm{h}}$ & $1.0$ eV \\
Crystal density & $\rho$ & 5360 kg/m$^{3}$   \\
Wave function widths: & & \\
\hspace{0.7em} electron in-plane & $l_{\mathrm{e}}$ & 4.4 nm  \\
\hspace{0.7em} hole in-plane & $l_{\mathrm{h}}$ & 3.6 nm  \\
\hspace{0.7em} electron/hole in $z$-direction & $l_{z}$ & 1.0 nm  \\
\hline
\end{tabular}
\caption{\label{tab:param}The GaAs material parameters and QD
system parameters used in the calculations (partly after Refs.
\onlinecite{adachi85,strauch90}).}
\end{table}

\section{Polarization dephasing after a sequence of ultrashort pulses}
\label{sec:delta}

The natural starting point for the discussion of the
multiple-pulse control of the optical polarization in a QD seems
to be the limit of ultrashort pulses. In this case, each pulse in
the sequence acts on a time scale much shorter than the lattice
response times. The subsequent evolution corresponds to a
relaxation of the lattice to a new equilibrium that is determined
by the optically created confined charge distribution. During this
process the exciton occupation cannot change but the
carrier-phonon correlations that appear in the system partly
destroy the exciton coherence. The resulting decay of the optical
polarization may be described analytically, both for a single
pulse \cite{krummheuer02} and for a sequence of pulses
\cite{vagov02a}.

It has been shown experimentally\cite{borri01,borri03} that the
optical polarization excited by an ultra-short pulse undergoes an
initial dephasing on a picosecond time scale, followed by a much
slower exponential decay. Model calculations  reveal that the
initial decay in these systems can be mostly attributed to LA
phonons \cite{vagov03,vagov04}. In general, decoherence refers to
the loss of definite phase relations between various components of
a quantum superposition. In our case these components are the
crystal ground state and the QD exciton state. Most easily
affected by decoherence is the superposition state with equal
amplitudes, which is reached from the ground state by an
$\alpha=\pi/2$ rotation on the Bloch sphere. In the following we
will concentrate on minimizing the decoherence occurring after
such a $\pi/2$ rotation. Other optimization goals have been
considered and treated with other strategies \cite{hohenester04}.

\begin{figure}[tb]
\begin{center}
\unitlength 1mm
\begin{picture}(85,30)(0,5)
\put(0,0){\resizebox{85mm}{!}{\includegraphics{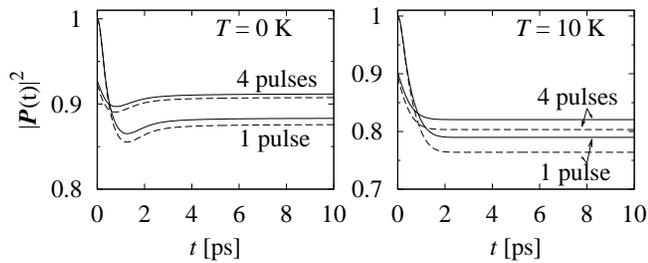}}}
\end{picture}
\end{center}
\caption{\label{fig:exact}The evolution of the optical
polarization excited with one or four infinitely short pulses.
Solid: exact result, dashed: perturbative result.}
\end{figure}

Analytical results for LA phonon-induced pure dephasing obtained
along the lines described in Ref.~\onlinecite{vagov02a} are
plotted in Fig.~\ref{fig:exact}. Shown is the evolution after a
single $\pi/2$ pulse and after a series of four identical,
in-phase, equidistant $\pi/8$ pulses. The total length of the
series is $t_{\mathrm{tot}}=1$ ps and the last pulse arrives at $t=0$. The
spectral cut-off resulting in real experiments from the finite
pulse duration is introduced in this $\delta$-pulse model by
including only the acoustic phonon response, which amounts to the
assumption that the actual pulse duration is still longer than LO
phonon periods and also long enough to spectrally avoid higher
excitations.

Compared to the single-pulse excitation, the initial polarization
in the 4-pulse case is reduced due to dephasing in between the
pulses. However, the final level of polarization is higher. This
means that the multiple-pulse excitation leads to a lower
contribution of the phonon-assisted transitions in the final
state. At zero temperature the loss of coherence compared to the
single-pulse initial value is reduced from about 12\% to about
9\%, i.e., we obtain an improvement of 25\%. In the remaining part
of the paper we will explore the possibilities of reducing the
degree of dephasing, going beyond the $\delta$-pulse limit and,
finally, extending the modeling to higher-energetic excitations.

\section{Phonon-induced perturbation of the coherent dynamics}
\label{sec:formal}

It may be expected that the reduction of dephasing described in
the previous Section may be extended by increasing the number of
pulses and optimizing their amplitudes. However, the analytical
formulas become highly involved for growing numbers of pulses. In
any case the spectral width of a sequence of $\delta$ pulses will
be infinite (although, possibly, the spectra may get non-trivially
shaped). Here, it should be recalled that the spectral selectivity
of real experiments has been accounted for in this approach by
concentrating on the lattice response in the low-frequency regime
and disregarding in the material model all higher energetic
excitations. A more accurate treatment of the spectral selectivity
is, of course, to introduce a spectral cut-off. This requires
using finite-length pulses; a case for which within our model no
exact solution of the dynamics  is known.

In order to deal with these issues
and to continue and extend our discussion without
restricting it to linear optical effects we resort to a scheme
which is non-perturbative in the electromagnetic field but
includes the phonon effects only in the leading order.

To calculate the phonon-related perturbation we expand
the formal expression for the density matrix of the total system
(in the interaction picture) at the final time $t$ up to the
second order in the phonon coupling constants \cite{cohen98}.
\begin{eqnarray}\label{evol0}
    \tilde{\varrho}(t) & = & \tilde{\varrho}(t_{\mathrm{i}})
    +\frac{1}{i\hbar}\int_{t_{\mathrm{i}}}^{t}d\tau
        [H_{\mathrm{int}}(\tau),\varrho(t_{\mathrm{i}})] \\
 & & -\frac{1}{\hbar^{2}}\int_{t_{\mathrm{i}}}^{t}d\tau
    \int_{t_{\mathrm{i}}}^{\tau}d\tau'
      [H_{\mathrm{int}}(\tau),
        [H_{\mathrm{int}}(\tau'),\varrho(t_{\mathrm{i}})]], \nonumber
\end{eqnarray}
where $\tilde{\varrho}(t)$ and $H_{\mathrm{int}}(\tau)$ are,
respectively, the density matrix of the total system and the
carrier-phonon interaction Hamiltonian in the interaction picture
with respect to the laser-induced evolution (which, in the case of
resonant coupling, is known exactly) and $t_{\mathrm{i}}$ is the
initial time. We assume an uncorrelated initial state with the
lattice in thermal equilibrium and the exciton subsystem in the
ground state. This procedure does not involve any Markovian
approximations; the truncation at the second order is valid as
long as the correction to the density matrix remains small.

After tracing out the lattice degrees of freedom and setting
$\alpha=\pi/2$ the polarization at the time $t$, which is
proportional to $\varrho_{01}(t)$, may be written as
\begin{equation}
\label{polar}
    |\bm{P}(t)|^{2}= |\bm{P_{0}}|^{2}\left[
        1-\int d\omega \frac{R(\omega)}{\omega^{2}}
            \left|K(\omega)-e^{i\omega t} \right|^{2}
                \right],
\end{equation}
where
\begin{equation}
\label{K}
    K(\omega)=\int_{-\infty}^{\infty}d\tau e^{i\omega \tau}
        \frac{d}{d\tau}\sin\Phi(\tau),
\end{equation}
\begin{displaymath}
        \Phi(\tau)=\int_{-\infty}^{\tau}d\tau' f(\tau').
\end{displaymath}
The upper limit of integration in Eq.~(\ref{K}) has been extended
to infinity under the assumption that at time $t$ the pulse has
already been completely switched off (the details of the formalism
are described in Ref.~\onlinecite{grodecka04a}). Since the
function $R(\omega)/\omega^{2}$ is regular at $\omega=0$, the
asymptotic limit of the polarization at $t\to\infty$ is (cf.
Ref.~\onlinecite{krummheuer02})
\begin{equation}
\label{asympt}
    |\bm{P}(\infty)|^{2}= |\bm{P_{0}}|^{2}\left[
        1-\int d\omega \frac{R(\omega)}{\omega^{2}}
            \left(|K(\omega)|^{2}+1 \right)
                \right].
\end{equation}

The phonon spectral densities $R(\omega)$ characterize the
spectral properties of the lattice and are given by
\begin{eqnarray}\label{spdens-expli}
    R(\omega) &  = & \frac{1}{\hbar^{2}}
    |n_{\mathrm{B}}(\omega)+1| \\
 & & \times
    \sum_{\bm{k}}\left|F_{\bm{k}}\right|^{2}
    \left[\delta(\omega-\omega_{\bm{k}})
    +\delta(\omega+\omega_{\bm{k}}) \right],\nonumber
\end{eqnarray}
where $n_{\mathrm{B}}(\omega)$ is the Bose distribution function.

It should be noted that part of the polarization drop can be
attributed to the deformation polaron formation around the exciton
(phonon dressing) which is not necessarily irreversible. In fact,
a perfectly adiabatic, thus reversible, passage to the
exciton--no-exciton superposition corresponds to vanishing
$K(\omega)$ and, although no irreversible dynamics takes place,
there is still a relative polarization reduction
\begin{displaymath}
    \delta_{\infty}=\int d\omega \frac{R(\omega)}{\omega^{2}},
\end{displaymath}
which corresponds to 1/2 of the asymptotic value after a $\delta$
pulse.

\begin{figure}[tb]
\begin{center}
\unitlength 1mm
\begin{picture}(85,87)(0,5)
\put(0,0){\resizebox{85mm}{!}{\includegraphics{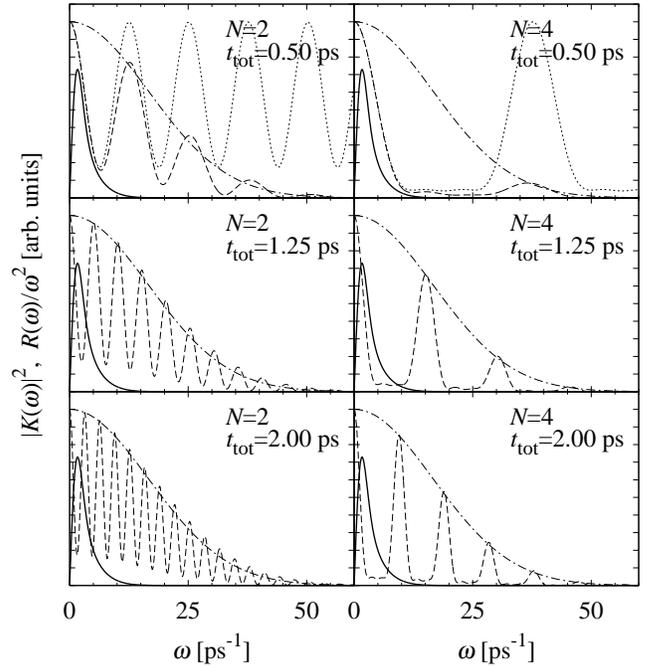}}}
\end{picture}
\end{center}
\caption{\label{fig:sw-general}Spectral characteristics of the
phonon reservoir and of the driven dynamics. Solid line: The
spectral density $R(\omega)/\omega^{2}$ at $T=0$ K for LA phonons.
Dashed line: the nonlinear spectrum $|K(\omega)|^{2}$ for a
sequence of identical short pulses for $\alpha=\pi/2$ and pulse
numbers and total durations as shown. Dash-dotted line: the
envelope calculated as the \emph{linear} spectrum of a single
pulse (up to an appropriate factor). In the upper figures also the
nonlinear spectrum for a sequence of $\delta$-pulses is shown
(dotted).}
\end{figure}

Since the spectral function $K(\omega)$ [Eq. (\ref{K})] is a
nonlinear functional of the pulse envelope $f(t)$, the only way of
calculating it is, in general, to numerically perform the Fourier
integral. However, some insight in its structure is possible in
the case of a series of non-overlapping control pulses. Let us
denote the $n$th pulse envelope ($n=1,\ldots, N$) by
$f_{n}(t-t_{n})$, where $t_{n}$ is the time at which the pulse is
applied, and the value of $\Phi(t)$ after the $n$th pulse by
$\alpha_{n}$, with $\alpha_{0}=0$ and $\alpha_{N}\equiv\alpha$.
Then, during the $n$th pulse one has
\begin{displaymath}
    \Phi(\tau)=\alpha_{n-1}+\tilde{\Phi}_{n}(\tau-t_{n}),
\end{displaymath}
where
\begin{displaymath}
    \tilde{\Phi}_{n}(\tau)=\int_{-\infty}^{\tau}f_{n}(\tau')d\tau'.
\end{displaymath}
Substituting this into Eq.~(\ref{K}) one finds
\begin{eqnarray}\label{K-grid}
    K(\omega) & = &\frac{1}{2i}\left[ F(\omega)- F^{*}(-\omega) \right], \\
\label{F-grid}
    F(\omega) & = & \sum_{n=1}^{N}\tilde{F}_{n}(\omega)
        e^{i(\alpha_{n-1}+\omega t_{n})},
\end{eqnarray}
with
\begin{displaymath}
    \tilde{F}_{n}(\omega)=\int dt e^{i\omega t}
        \frac{d}{dt}e^{i\tilde{\Phi}_{n}(t)},
\end{displaymath}
Since, by definition, the function $\tilde{\Phi}_{n}(t)$ changes
only around $t=0$, the transform $\tilde{F}_{n}(\omega)$ is a
smooth envelope of width inversely proportional to the duration of
the pulse $f_{n}(t)$. If the rotation angle per pulse is small
this envelope function is close to the linear Fourier transform of
$f_{n}(t)$ (up to a factor).

In the limit of infinitely short pulses the transform
$\tilde{F}_{n}(\omega)$ becomes infinitely broad and one has
\begin{equation}\label{Fn}
    \tilde{F}_{n}(\omega)=\tilde{F}_{n}(0)=
        e^{i(\alpha_{n}-\alpha_{n-1})}-1.
\end{equation}
If, in addition, the pulses are equally spaced, $t_{n}=n\Delta t$,
then the resulting spectral function $F(\omega)$ is periodic in
the frequency domain with the period $2\pi/\Delta t$. This limit
corresponds to the exact solution obtained previously and allows
one to compare both results and estimate the error introduced by
the perturbative approach. As can be seen in Fig.~\ref{fig:exact},
the perturbative results not only reproduce qualitatively all the
essential features of the exact result but are also reasonably
close to the exact results, at least at low temperatures.

Temporally broadening the pulses results in the appearance of the
envelopes $\tilde{F}_{n}(\omega)$ [Eq.~(\ref{F-grid})] which break
the periodicity and eliminate the high-frequency part of the
nonlinear spectrum. For a series of identical, weakly overlapping,
equally spaced pulses one has $\tilde{F}_{n}(\omega)\equiv
\tilde{F}(\omega)$, $\alpha_{n}=n\alpha /N$, and
$t_{n}=(n-1)\tilde{T}/N$. In order to simplify the formulas, we define
here $\tilde{T}=Nt_{\mathrm{tot}}/(N-1)$, where $t_{\mathrm{tot}}$ is
the total length of the pulse sequence measured from the center of the
first pulse to the center of the last. 
The summation in Eq.~(\ref{F-grid}) may be
easily performed analytically and the result for $\alpha=\pi/2$ is
\begin{eqnarray}\label{diffract}
    \lefteqn{| K(\omega) |^{2} =} \\
\nonumber
 & & \frac{1}{4}|\tilde{F}(\omega)|^{2}\left[
    \frac{\sin^{2}
     \frac{\tilde{T}}{2}(\omega+\frac{\pi}{2\tilde{T}})}%
        {\sin^{2}
            \frac{\tilde{T}}{2N}(\omega+\frac{\pi}{2\tilde{T}})}
   + \frac{\sin^{2}
     \frac{\tilde{T}}{2}(\omega-\frac{\pi}{2\tilde{T}})}%
        {\sin^{2}
            \frac{\tilde{T}}{2N}(\omega-\frac{\pi}{2\tilde{T}})}\right].
\end{eqnarray}
This is formally identical to the sum of two diffraction patterns
for equally spaced optical grids of length $\tilde{T}$, modulated
by the envelope $\tilde{F}(\omega)$ which suppresses higher order
``diffraction maxima'', and shifted oppositely by
$\pm\pi/(2\tilde{T})$. This behavior is illustrated in
Fig.~\ref{fig:sw-general}, where we have chosen pulses with a
duration of $\tau_{p}=100$ fs (full width at half maximum, FWHM).
As seen in the figure, this choice assures that the relevant
spectral function $|K(\omega)|^{2}$ has a spectral cutoff below
the LO phonon frequency $\Omega=54$ ps$^{-1}$, consistent with the
neglect of this excitation in our model.

In this case of identical, weakly overlapping, equally spaced
pulses, for increasing pulse number the period of the spectrum in
the frequency domain increases, i.e. the higher maxima are shifted
towards higher frequency and decrease due to the envelope.
However, the width of the central maximum is determined by the
total duration of the sequence. From Eq. (\ref{diffract}) it can
be seen that when taking $\tilde{T}$ as a measure of the sequence
duration then this width is essentially insensitive to the pulse
number. Similarly, the value at $\omega=0$ is fixed by the gate
angle and is always $|K(0)|^{2}=\sin\alpha=1$. Modifying the total
duration of the sequence leads to proportional scaling of the
$\omega$-dependence of the nonlinear spectrum $|K(\omega)|^{2}$
(the ``diffraction pattern'') under the envelope. The latter is
fully determined by the single pulse spectrum which, for
$\alpha\lesssim\pi/2$ is very close to the linear one even for two
pulses (see Fig.~\ref{fig:sw-general}).

The independent control over the nonlinear spectrum for high
frequencies (by pulse width) and in the LA frequency domain (by
the sequence length and pulse number) opens the possibility of
optimizing the coherence by reducing the overlap in Eq.
(\ref{asympt}), alternative to using long smooth pulses. In the
next Section we study this issue, releasing the restriction to
equal pulses in order to find optimized pulse sequences.

\section{Reduction of dephasing by optimized pulse sequences}
\label{sec:sequence}

The aim of this Section is to quantify the reduction of the
LA-phonon induced dephasing of the optical polarization by using
optimized sequences of pulses with a fixed finite pulse length. As
discussed in the previous Section, by using pulses of finite
duration we impose an envelope on the nonlinear spectra of the
controlled dynamics and thereby assure that higher-frequency
spectral features, including LO phonons, higher exciton states and
confined two-pair excitations, have negligible impact.

In real systems, apart from the pure dephasing resulting from the
phonon response to the evolution of the exciton state there is
another contribution to the loss of fidelity related to
exponential decay processes (finite exciton lifetime in our
specific case) \cite{borri01,birkedal01}. The effect of the latter
on the polarization magnitude grows with the duration of the pulse
sequence. Therefore, a question of practical importance is to what
extent the system coherence may be preserved against the pure
dephasing during an operation performed by a sequence of a fixed
number of pulses within a definite time interval that is limited
by the exponential decoherence.

\begin{figure}[tb]
\begin{center}
\unitlength 1mm
\begin{picture}(85,30)(0,5)
\put(0,0){\resizebox{85mm}{!}{\includegraphics{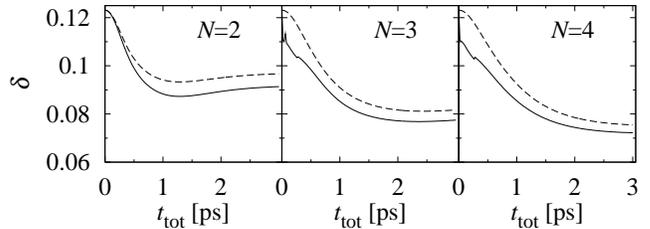}}}
\end{picture}
\end{center}
\caption{\label{fig:err234}The relative polarization drop (with
respect to the unperturbed case) for the $\alpha=\pi/2$ excitation
performed by sequences of $N=2,3,4$ equidistant narrow pulses
with $\tau_{\mathrm{p}}=100$ fs within the time interval $t_{\mathrm{tot}}$
at $T=0$ K for optimized pulse intensities (solid) and for identical
pulses (dashed).}
\end{figure}

By numerically calculating the nonlinear spectra $K(\omega)$ [Eq.
(\ref{K})] for sequences of $N$ equidistant narrow (but not
necessarily non-overlapping) pulses and maximizing the asymptotic
polarization magnitude [Eq.~(\ref{asympt})] with respect to the
pulse amplitudes for a fixed total rotation angle $\alpha=\pi/2$
we found optimized\footnote{We used the standard multi-dimensional
optimization procedure from the IMSL library.} 
pulse sequences for a given total duration of the
sequence $t_{\mathrm{tot}}$ and the resulting minimal polarization drop
\begin{displaymath}
    \delta=
    \frac{|\bm{P}_{0}|^{2}-|\bm{P}(\infty)|^{2}}{|\bm{P}_{0}|^{2}}.
\end{displaymath}
The result is shown in Fig.~\ref{fig:err234}.

As expected, it turns out that the degree of dephasing typically
tends to decrease as the time interval becomes longer. Optimizing
the pulse sequence leads to some further reduction of the
polarization drop compared to the case of equal pulses, although
this additional gain is relatively small. An interesting feature
is the slight increase of the decoherence drop for times
$t_{\mathrm{tot}}\gtrsim 1$ ps in the $N=2$ case. This may be understood with
the help of Fig.~\ref{fig:sw-general} by noting that for
$2\pi/t_{\mathrm{tot}}\simeq \omega_{\mathrm{LA}}$, where
$\omega_{\mathrm{LA}}$ is the LA spectral density cut-off, the
maximum of $R(\omega)/\omega^{2}$ coincides with the minimum of
$|K(\omega)|^{2}$ while for longer times the latter is quickly
oscillating and averages to its mean value.

For $t_{\mathrm{tot}}\lesssim\tau_{\mathrm{p}}$ the duration of the pulse
sequence is actually determined by $\tau_{\mathrm{p}}$
and $t_{\mathrm{tot}}/N$ becomes the shortest relevant time scale. 
Thus, the high frequency part of the pulse spectrum is governed by
$t_{\mathrm{tot}}/N$ in this case
and the simple discussion presented above for
non-overlapping pulses is not valid. 
In this case, the nonlinear spectra extend to
much larger frequencies and the overall error drops down due to
strong modulations of the resulting pulse shape. Although the
spectral cut-off cannot be assured \emph{a priori} in this case,
it is still possible to control the consistency of the calculation
by inspection of the nonlinear spectrum corresponding to the
resulting optimized pulse sequence.
In fact, it turns out that for very short $t_{\mathrm{tot}}$ 
the error values shown in Fig.~\ref{fig:err234} do not correspond to
pulse sequences with properly restricted spectra. Moreover, due to the
very complex structure of the error as a function of the pulse intensities
it is not guaranteed that the optimized value found by our procedure
in the strong overlap regime corresponds to the global minimum. 
For weak and moderate overlap the optimized sequences are likely to correspond to the
global minimum.

\begin{figure}[tb]
\begin{center}
\unitlength 1mm
\begin{picture}(85,55)(0,5)
\put(0,0){\resizebox{85mm}{!}{\includegraphics{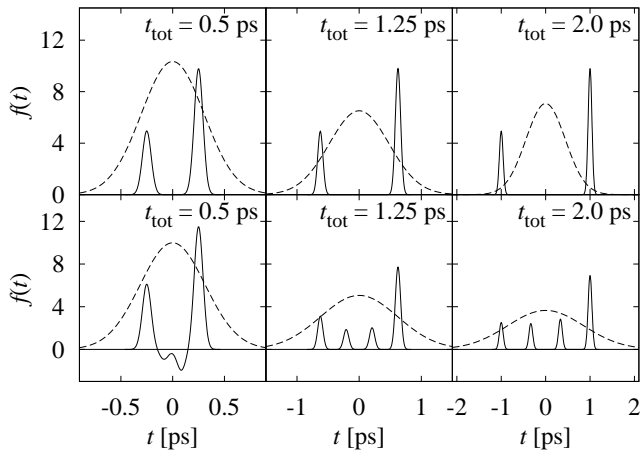}}}
\end{picture}
\end{center}
\caption{\label{fig:pulse24}The optimized sequences of $N=2$ (upper)
and $N=4$ (lower) narrow pulses ($\tau_{\mathrm{p}}=100$ fs)
compared to the Gaussian pulses (dashed, arbitrary scale)
yielding the same asymptotic polarization (at $T=0$ K).}
\end{figure}

It is interesting to compare the coherence loss for an excitation
performed by a sequence of pulses and by a single Gaussian pulse.
In Fig.~\ref{fig:pulse24} we compare the optimized pulse sequence
with a Gaussian pulse performing the same $\alpha=\pi/2$ rotation
with the same asymptotic polarization drop as obtained from the
sequence. It turns out that already for $N=2$ the pulse sequence
is shorter than the corresponding Gaussian pulse as long as
$t_{\mathrm{tot}}\lesssim 2$ ps. For very short pulse durations the Gaussian
may even be twice as long as the pulse sequence. For $t_{\mathrm{tot}}\gtrsim
2$ ps, the smooth Gaussian control can no longer be significantly
outperformed by a pulse sequence. However, as the number of pulses
is increased the pulsed control becomes more efficient over a
wider range of sequence durations.

Another interesting aspect worthwhile to note is that in the
infinitely short pulse limit the optimal sequence of $N=2$ pulses
does not depend on the structure of the spectral density of the
reservoir and for $\alpha=\pi/2$ it always corresponds to the
ratio of pulse areas of 1:2 (see the Appendix). As can be seen in
Fig.~\ref{fig:pulse24}, this universal area ratio of the optimal
2-pulse sequence holds approximately also for pulses of finite
duration.

\begin{figure}[tb]
\begin{center}
\unitlength 1mm
\begin{picture}(85,27)(0,5)
\put(0,0){\resizebox{85mm}{!}{\includegraphics{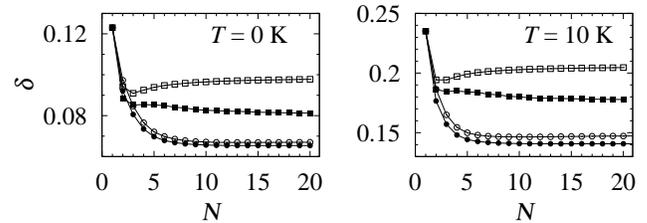}}}
\end{picture}
\end{center}
\caption{\label{fig:err-n}The polarization drop as a function of
the number of pulses $N$ for the total durations $t_{\mathrm{tot}}=1$ ps
(squares) and $t_{\mathrm{tot}}=5$ ps (circles) at $T=0$ K (left) and $T=10$ K
(right). Full symbols: optimized pulse intensities, empty symbols:
equal pulses. Here $\tau_{p}=100$ fs.}
\end{figure}

By increasing the number of pulses, the coherence loss may be
further reduced (Fig.~\ref{fig:err-n}). We find that at both
temperatures the polarization drop can be reduced by about 40\%.
Note that even in the adiabatic limit, 50\% of the polarization
drop is present. This implies that the irreversible effect is
reduced by about 80\%. The reduction strongly depends on the total
duration but in all cases (for $t_{\mathrm{tot}}$ in the picosecond range)
most of it is gained already for a few pulses. For short $t_{\mathrm{tot}}$
and for a control sequence consisting of equal pulses there is an
optimal number of pulses beyond which the error grows again. This
results directly from Eq.~(\ref{diffract}): for higher $N$ the
nonlinear spectrum decreases more slowly around $\omega=0$ which
determines the resulting overlap with $R(\omega)$ when the width
of the ``zeroth order'' maximum is comparable to the LA cut-off
frequency. This effect disappears if one allows variable pulse
intensities. 
From Fig.~\ref{fig:err-n} it is evident that 
for sufficiently long total durations $t_{\mathrm{tot}}$
of the pulse sequence the polarization drop $\delta$ 
comes close to its value in the adiabatic limit already when
the sequence comprises only a few pulses.
In this case, the optimazation of the intensities
yields only a marginal improvement as is seen from the
curves corresponding to $t_{\mathrm{tot}}=5$ ps 
in Fig.~\ref{fig:err-n}.
In contrast, for pulse sequences of short duration $t_{\mathrm{tot}}$,
$\delta$ is well above its adiabatic value, but
the improvement resulting from the
intensity optimization is now substantial
as illustrated by the $t_{\mathrm{tot}}=1$ ps 
curves in Fig.~\ref{fig:err-n}.

\begin{figure}[tb]
\begin{center}
\unitlength 1mm
\begin{picture}(85,40)(0,5)
\put(0,0){\resizebox{85mm}{!}{\includegraphics{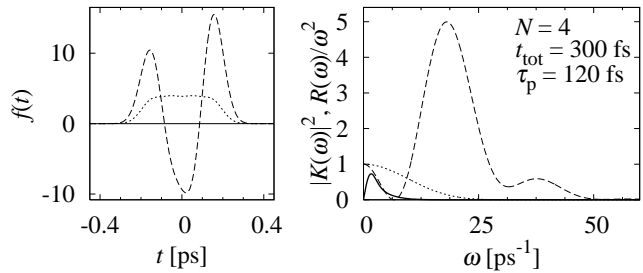}}}
\end{picture}
\end{center}
\caption{\label{fig:pushout}Left: the optimized pulse envelope
formed by the sequence of $N=4$ overlapping narrow pulses (dashed:
pulses with optimized amplitudes, dotted: equal pulses) for the
sequence duration $t_{\mathrm{tot}}=0.3$ ps at $T=0$ K. Right: the
corresponding nonlinear spectra $|K(\omega)|^{2}$ along with the
phonon spectral density (solid).}
\end{figure}

In order to see most clearly how the optimization of overlapping
pulses leads to a decrease of decoherence let us compare the
spectral features corresponding to equal and optimized pulses in
the case of strong overlap (Fig.~\ref{fig:pushout}). In the
optimized case, a strong modulation of the resulting pulse
envelope is possible which leads to considerable reduction of the
nonlinear spectrum in the LA sector. Such a modulation produces,
however, large spectral features beyond the LA cut-off, as can be
clearly seen in Fig.~\ref{fig:pushout}. In order to provide the
desirable spectral cutoff the pulse length must now be increased,
compared to the case of non-overlapping pulses, or further
constraints have to be introduced in the optimization procedure.

\section{Extending the model}
\label{sec:extend}

As discussed above, the pulse lengths used to control the quantum
state must be long enough to assure a frequency cut-off that
avoids in any case the undesired dynamics that otherwise may
result from high energetic excitations. This restriction may be
inconvenient for practical applications as it implies the use of
long pulses as building blocks of the control sequence. Below we
show how the optimization procedure may be extended such that the
undesired dynamics is not avoided by a frequency cut-off but
rather by suitably shaping the pulse sequence. To this end we
first need to extend our model and include explicitly the
pertinent higher energetic excitations.

\begin{figure}[tb]
\begin{center}
\unitlength 1mm
\begin{picture}(85,35)(0,5)
\put(0,0){\resizebox{85mm}{!}{\includegraphics{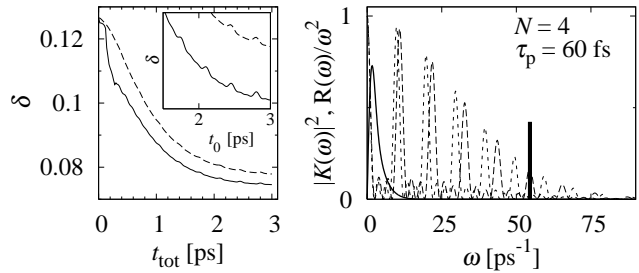}}}
\end{picture}
\end{center}
\caption{\label{fig:err4-LO}Left: the asymptotic polarization
decay induced by LA and LO phonons for $N=4$ pulses and
$\tau_{\mathrm{p}}=60$ fs; solid: optimized sequence, dashed:
equal pulses. The inset shows part of the curves in magnification.
Right: The nonlinear spectrum $|K(\omega)|^{2}$ corresponding to
two values of the pulse sequence duration $t_{\mathrm{tot}}=1.74$ ps (strong
overlap with the LO line; dashed) and $t_{\mathrm{tot}}=1.92$ ps (weak
overlap; dotted). Also shown are the spectral densities for LA
(solid) and LO (bar, arbitrary scale) phonons.}
\end{figure}

The first obvious step is to include LO phonons which, for
confined states, contribute a single line (due to their weak
dispersion around $k=0$). In the plot of the polarization drop
against the duration (Fig.~\ref{fig:err4-LO}, left) small
additional features appear periodically, both for equal as well as
for optimized pulses (visible in the inset). They result from the
fact that in the case of weakly overlapping pulses the positions
of the maxima in the nonlinear spectrum $|K(\omega)|^{2}$ are
fixed by the total duration $t_{\mathrm{tot}}$ and the total number of
pulses, and the decoherence effect increases each time one of them
overlaps with the LO phonon line (Fig.~\ref{fig:err4-LO}, right).
In the system under discussion this effect is rather weak because
of the strong reduction of the LO phonon line due to charge
cancellation. Note that extending the reservoir model allows us to
reduce the pulse duration to 60 fs which now provides the spectral
cutoff before the onset of higher-frequency reservoir features.

The next higher excitations of the phonon bath are related to the
possibility of phonon-assisted transitions to dark states
\cite{machnikowski04a} (which may be viewed as an effect of
nonadiabaticity of exciton dynamics with respect to LO lattice
oscillations\cite{fomin98}). 
We extend the Hamiltonian (\ref{ham}) by adding terms describing
the excited exciton states and their
LO-phonon-assisted coupling to the ground state
(other couplings do not contribute to the leading order),
\begin{eqnarray*}
\lefteqn{H_{\mathrm{exc}}=}\\
&&\sum_{n>1}E_{n}|n\rangle\!\langle n|
+\sum_{n>1}|1\rangle\!\langle n|\sum_{\bm{k}(\mathrm{LO})}
F^{(n)}_{\bm{k}}(b_{\bm{k}}+b_{-\bm{k}}^{\dag})+\mathrm{H.c.},
\end{eqnarray*}
where $E_{n}$ are the energies of the excited states and the coupling
constants $F^{(n)}_{\bm{k}}$ are analogous to those 
given by Eqs.~(\ref{F-LA}) and (\ref{F-LO}) and may be found e.g. in
Ref. \onlinecite{jacak03b}. This coupling induces additional
decoherence with the corresponding asymptotic polarization drop given by
\begin{eqnarray*}
\lefteqn{\Delta|\bm{P}(\infty)|^{2}=}\\
&&-|\bm{P}_{0}|^{2}\sum_{n>1}\int d\omega
\frac{R^{(n)}(\omega-E_{n})}{\omega^{2}}
\left(|L(\omega)|^{2}+\frac{1}{2}\right),
\end{eqnarray*}
where 
\begin{equation}
\label{K12}
    L(\omega)=\int_{-\infty}^{\infty}d\tau e^{i\omega \tau}
        \frac{d}{d\tau}\sin\frac{1}{2}\Phi(\tau),
\end{equation}
and $R^{(n)}(\omega)$ are defined as in Eq.~(\ref{spdens-expli}) but
with the coupling constants $F^{(n)}_{\bm{k}}$ (see also 
Ref.~\onlinecite{grodecka04a}). The optimal control, as measured by
the minimal polarization drop, corresponds now to minimizing the
sum of the overlap integral appearing in Eq.~(\ref{asympt}) and that
given above.

\begin{figure}[tb]
\begin{center}
\unitlength 1mm
\begin{picture}(85,35)(0,5)
\put(0,0){\resizebox{85mm}{!}{\includegraphics{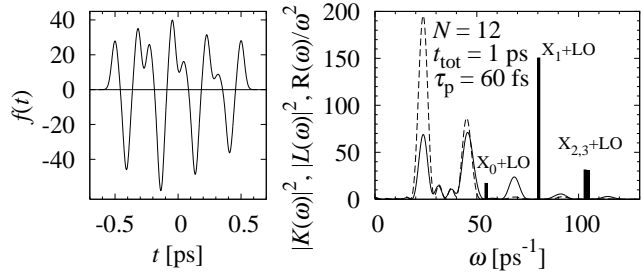}}}
\end{picture}
\end{center}
\caption{\label{fig:4states}The pulse sequence (left) and the
nonlinear spectra (right) corresponding to the optimal control in
the model extended by adding LO-phonon-assisted excitation of a
few lowest dark states (X$_{1}$--X$_{3}$). In the right panel, the
solid and dashed lines show the spectral functions $|K(\omega)|^{2}$
and $|L(\omega)|^{2}$, respectively, while the vertical bars represent
the spectral features related to  phonon-assisted transitions to 
optically inactive states.}
\end{figure}

For the results shown in Fig.~\ref{fig:4states}, we include three
optically inactive excited states at 17, 32 and 33 meV above the ground
state. The optimization
was done for a relatively high number of pulses, so that they
overlap giving rise to a single, strongly shaped envelope. The
simple ``diffraction pattern'' argument does not hold in this case
and strongly nonlinear effects become important. As a result,
additional flexibility appears in the shaping of the nonlinear
spectra and the pulse optimization leads to spectral functions
that to a large extent avoid the overlap with these discrete
transitions as can be seen in Fig.~\ref{fig:4states}.

\section{Conclusions}
\label{sec:conc}

We have shown that the coherence of the optically generated
superposition of exciton occupation states confined in a quantum
dot may be considerably increased by using a sequence of narrow
(but finite) pulses to drive the carrier system. By doing so, one
modifies the nonlinear spectral properties of the optically driven
evolution of the carrier subsystem which may decrease the overlap
between these nonlinear spectra and the spectral features of the
phonon reservoir.

Instead of expressing the admissible class of pulses in terms of
mathematically defined constraints, the proposed approach relies
on physically transparent semi-quantitative conditions. By using
pulses of finite duration as building blocks one controls the
frequency cut-off so that the spectra of the resulting pulses may
be flexibly restricted to the frequency range of which one has
reliable knowledge. In this way we guarantee that the result of
the optimization is defined by the physical content of the
carrier-phonon model and not by its unphysical truncations. On the
other hand, by optimizing the pulse amplitudes the spectral
properties of the resulting evolution are chosen in such a way
that resonances with phonon modes are minimized.

Given a finite control time window, it is not possible to
completely avoid the excitation of phonon modes and the resulting
dephasing of the exciton states (measurable via the decay of the
optical polarization). This is due to the fact that the whole
continuum of phonon frequencies cannot be avoided simultaneously.
Nevertheless, with an optimized choice of pulse intensities the
degree of dephasing may be considerably reduced, both at zero and
finite temperatures, to a degree depending on the allowed duration
of the pulse sequence (as expected, longer control time windows
admit larger reduction of dephasing).

From the practical point of view, the results obtained in this
paper are remarkable with two respects. First, in the picosecond
range of control times (relevant for phonon-induced pure dephasing
effects) using sequences of a few (2--4) pulses leads to higher
final degree of coherence of the resulting state than that
obtained by a single Gaussian pulse of the same total duration. 
This may be important since obtaining
short series of phase-locked pulses is certainly much more
feasible than generating an arbitrarily shaped pulse resulting
from a full optimization procedure.

Second, it is found that increasing the pulse number beyond 3 or 4
brings only small improvements, compared to that gained by
replacing the single pulse by a series of 2 or 3. This shows that
the limited possibility of reducing the dephasing may be almost
fully exploited already with limited control resources. The
universality of the optimal sequence of two pulses further
increases the usefulness of these results.

\begin{acknowledgments}
We gratefully acknowledge the financial support for P.M. by the
Alexander von Humboldt Stiftung. The work was partly supported by
the Polish KBN Grant No. PB 2 PO3B 085 25.
\end{acknowledgments}

\section*{Appendix}
\begin{appendix}
\section*{Universally optimal 2-pulse sequence}

In this appendix we show that the optimal amplitudes of two
ultrashort pulses (in the limit of infinitely short pulses) depend
only on the total rotation angle but not on the properties of the
lattice reservoir.

Let us denote the total rotation angle by $\alpha$ and the
rotation angles performed by the first and second pulses
(proportional to the corresponding pulse areas) by $\alpha_{1}$
and $\alpha_{2}=\alpha-\alpha_{1}$, respectively. Assuming the
pulses to arrive at $t=\pm t_{\mathrm{tot}}/2$ and inserting Eq.~(\ref{Fn})
into Eqs.~(\ref{F-grid}) and (\ref{K-grid}) one finds
\begin{eqnarray*}
    \lefteqn{\left| K(\omega) \right|^{2}=} \\
& &  \sin^{2}\alpha+2\left(1-\cos\frac{\omega t_{\mathrm{tot}}}{2}\right)
        \left( \sin^{2}\alpha_{1}-\sin\alpha\sin\alpha_{1}
        \right).
\end{eqnarray*}
Since $1-\cos\omega t\ge 0$, the minimum (with respect to
$\alpha_{1}$) of the overlap with the nonnegative function
$R(\omega)/\omega^{2}$ [Eq.~(\ref{asympt})] always corresponds to
the minimum of the rightmost bracket in the above formula.
Therefore, irrespective of the specific form of $R(\omega)$, the
optimal choice is
\begin{displaymath}
    \sin\alpha_{1}=\frac{1}{2}\sin\alpha.
\end{displaymath}
For $\alpha=\pi/2$ this corresponds to $\alpha_{1}=\pi/6$ and
$\alpha_{2}=\pi/3$, so that the pulse area ratio is 1:2. This
result holds, however, only in the leading order in the phonon
coupling and is no longer valid in the general, non-perturbative
case.
\end{appendix}


\end{document}